Big data challenge for social sciences: from society and opinion to replications*

**Big data challenges for the social sciences: from society and opinion to replications**

**By Dominique Boullier**
**CDH, Social Media Lab, EPFL Lausanne**

**May 2016**

*Dominique Boullier, EPFL, CDH, Social Media Lab
email:dominique.boullier@epfl.ch, tel: 0033 (0) 608538560

# Big data challenge for social sciences: from society and opinion to replications

## Abstract

Big Data dealing with the social produce predictive correlations for the benefit of brands and web platforms. Beyond "society" and "opinion" for which the text lays out a genealogy, appear the "traces" that must be theorized as "replications" by the social sciences in order to reap the benefits of the uncertain status of entities' widespread traceability. High frequency replications as a collective phenomenon did exist before the digital networks emergence but now they leave traces that can be computed. The third generation of Social Sciences currently emerging must assume the specific nature of the world of data created by digital networks, without reducing them to the categories of the sciences of "society" or "opinion". Examples from recent works on Twitter and other digital corpora show how the search for structural effects or market-style trade-offs are prevalent even though insights about propagation, virality and memetics could help build a new theoretical framework.

When in 2007 Savage and Burrows pointed out "the coming crisis of empirical methods", they were not expecting to be so right. Their paper however became a landmark signifying the social sciences' reaction to the tremendous shock triggered by digital methods. As they frankly acknowledge in a more recent paper, they did not even imagine the extent to which their prediction might become true, in an age of Big Data, where sources and models have to be revised in the light of extended computing power and radically innovative mathematical approaches. They signalled not just a debate about academic methods but also a momentum for "commercial sociology" in which platforms acquire the capacity to add "another major nail in the coffin of academic sociology claims to jurisdiction over knowledge of the social", because "research methods (are) an intrinsic feature of contemporary capitalist organizations" (Burrows and Savage, 2014). This need for a serious account of research methods is well tuned with the claims of Social Studies of Science that should be applied to the social sciences as well.

I would like to build on these insights and principles of Burrows and Savage to propose an historical and systematic account of quantification during the last century, following in the footsteps of Alain Desrosières, and in which we see Big Data and Machine Learning as a major shift in the way social science can be performed. And since, according to Burrows and Savage (2014), "the use of new data sources involves a contestation over the social itself", I will take the risk here of identifying and defining the entities that are supposed to encapsulate the social for each kind of method: beyond the reign of "society" and "opinion", I will point at the emergence of the "replications" that are fabricated by digital platforms but are radically different from previous entities. This is a challenge to invent not only new methods but also a new process of reflexivity for societies, made available by new stakeholders (namely, the digital platforms) which transform reflexivity into reactivity (as operational quantifiers always tend to). This great transformation is built by and for the sake of brands which are framing everyone's perceptions, from truly commercial brands to scholars for their H-index, cities for their attractiveness, youngsters on YouTube or politicians on Twitter. This is why "the end of theory" (Anderson, 2008) concerns everybody. It also means the end of some kind of reflexivity, the end of the opportunity to connect long-term trends and opinion movements to personal experience by focusing only on replications, the high frequency dimension of the social, although the social is made of all three of these "wavelengths" (Society, Opinion, Replications). My proposal is not to disqualify data and processes related to high-frequency replications, but to maintain a pluralistic range of analysis while agreeing to take up the challenge of Big Data and Machine Learning. Finally, there is no reason why the social sciences' "authority" should escape the challenge to all kinds of authority, intermediation and power generated by the digital revolution. This seems more compatible with a pragmatist approach, which requires us to understand the social from within while not being trapped in the hype of the promotion of this new "apparatus" that is yet to be established as a convention (Eymard-Duvernay et al., 2004).

This is a way of accounting for the "social life of methods" (Law, Ruppert and Savage, 2011), even though that might become painful for social sciences, as Geoff Bowker said: "For those of us brought up learning that correlation is not causation, there's a certain reluctance to examine the possibility that correlation is basically good enough. It is surely the case that we are moving from the knowledge/power nexus portrayed by Foucault to a data/action nexus that does not need to move through theory: all it needs is data together with preferred outcomes" (Bowker, 2014). This concern however regards the Machine Learning approach more than it does Big Data as such, which is clearly too large a term for very different processes. This is why I shall divide the issues addressed here into two parts:

1/ those related to the very nature of "Data", where we emphasize: a) their granularity more than their size; b) their target on social behaviours more than the general use of large data bases as it exists in many scientific and technical activities[1]; and c) their velocity, which is a very distinctive feature for the social sciences, even for those familiar with large databases such as censuses and registries of any kind (for instance patents databases). We should speak more of "Fast Data" than Big Data, and this is specifically the feature that requires a new generation of social sciences.

2/ those related to Machine Learning and its inductive method. Of course they rely on the qualities of Big Data sources to make the learning algorithms work. But the "data-to-model" approach is still very challenging, even for computer scientists who stick to "model-to-data". Many projects and systems that look like they are using Big Data use neither inductive methods nor data-to-model approaches, and make an extensive use of ontologies for instance, which do not require Machine Learning algorithms, although they do use classification methods on a large scale.

Therefore, the main challenges for the social sciences are twofold: fast data, and inductive methods whenever they are used for tracing social behaviours (and not in genomics or in astrophysics for instance). This framework having been set up, I shall discuss each of these challenges at the end of the paper after having compared them to other "epochs" of quantification, in order to understand the conditions required for these new conventions to be built.

## 1. Trouble over entities in the Digital Age

***Neither people nor identities nor communities but traces are the "raw" material***

For many years, but in an extended way with social networks, computer science has calculated and modelled the social as if the traces collected allowed access to the "truth" about individuals in a more effective way than do polls, surveys and censuses. Consider two examples, one academic and the other commercial:
- *"The Web does not just connect machines, it connects people."* (Knight Foundation, 14 September 2008). There you are, this is what Sir Tim Berners-Lee, co-founder of the Web in 1991, with René Caillau, stated to emphasize the transition to a dimension of networks which is neither technological (III for International Information Infrastructure) nor documentary (WWW), but social (GGG for Global Giant Graph).
- Facebook has managed the tour de force of "normalising" members' declaration of their true identity, that is to say, the features of identity provided by the civil registry, the name and surname, as opposed to the tradition of anonymity on the Web. The platform thus claims to have become the authority of reference or even a civil-registry-alternative, competing with Google in this regard. This is becoming usual whenever one uses a Facebook account for creating and certifying access to other apps, for instance.

---

[1] This is what is amazing about the hype of Twitter studies from computer scientists who show no concern for the traditions of social sciences network analysis, but who would not dare to act so casually with data bases from the hard sciences. As H. Wallach put it : *"Few computer scientists or engineers would consider developing models or tools for analyzing astronomy data without involving astronomers. So, why, then, are so many methods for analyzing social data developed without the involvement of social scientists? I think, in part, that it is because, as humans, we have strong intuitions about the social world."*
([Big Data, Machine Learning, and the Social Sciences: Fairness, Accountability, and Transparency.](#) H. Wallach. Available on the Medium website, 2014.)

Yet there is no guarantee whatsoever of any connection between the identities on Facebook, or Berners-Lee's "people", and persons identified by their civil registry. What are connected are merely the retrieved accounts and data, and these are only the traces of activity from an entity, which could possibly take on the form of a legal civil status. Based on the scores that classify sites on a search engine, the resulting topology of sites and blogs never discusses their contents as such, but rather the inbound and outbound links that produce a rank of authority or hub, as defined in the network topology (Kleinberg et al, 1998) and are not a civil status. It should be noted here from the outset what I mean by traces, in order to distinguish them from data. Traces can range from signals (so-called "raw" ones, generated by objects) to unstructured verbatim; they can be traces exploited in databases (links, clicks, likes, cookies)[2] by operators or platforms, but also captured independently of this through the API and, as such, fall outside permanent relational databases (Bowker and Star, 1999). Traces are not necessarily pre-formatted for a specific calculation nor are they dependent upon aggregation that can then be applied. It is easy to argue that, despite everything, "behind" these sites or "behind" these clicks, there are most certainly people, but that does not alter the fact that the algorithms themselves do not take this onto consideration and that, furthermore, no guarantees can be given in this regard. Traces understood in the restricted sense are produced by platforms and digital-technological-systems, but are not the "signs" or evidence of anything other than themselves, as long as relationships with other attributes are not created and validated. Of course, they are transactional data because there is no trace without a distributed setting of relationships including humans and non-humans. This differs radically from the data that can be recovered en masse from client files or from administrative acts. Certainly, the Big Data methods for calculating can be applied here in both cases, but the traces are a priori independent of other attributes, in particular socio-demographic features, which are rarely used in correlations between traces. Relationships with more conventional parameters in data sciences are limited to time (a timestamp) and location (geo-location tags), which allow for the production of timelines and maps that become simplified modes of representation for traces. Can the social sciences accept this shift in the "raw" material they usually process?

### *Traces are produced by platforms*

In order to get into a thicker description of this social life of data, consider how it works for Amazon or Apple. The Web is no longer distributed but monopolized by these four platforms – GAFA that centralize the majority of traffic, with Twitter extending this traces industry and IBM focusing more on the "big stuff"). What I described above about traces and their detachment from the legal ID's features was more central for platforms such as Google and Facebook. The pretention of these platforms to perform their version of a "society" should be noticed according to the three wavelengths of social sciences above mentioned.. On such consumer platforms as Amazon and Apple, it is not people who are linked to one another but above all tastes (books or music originally), reflected in traces of purchases, and thus of choices, which can be treated en masse to produce patterns and profiles, independently of personal information.  It should certainly not be forgotten that all these platforms without exception are also very fond of civil status-type data, phone numbers and other highly attractive resources to advertisers, to whom they are sold.  Amazon and Apple are designing the perfect market environment in which preferences can be traced, they built their own sociological device focused on "opinion", the second wavelength I mentioned above. The marketing methods thus developed are largely based on mass advertising or on emails addressed to IP addresses, or emails that have clicked on an article (retargeting), and much more rarely on sophisticated links with other attributes of the supposed people attached to these addresses or clicks (profiling). Traces of digital behaviour are thus a particularly profitable "raw material", without the need to appeal to the social sciences, although platforms, social listening agencies and marketing experts readily make use of academically coined terms such as "communities", "networks", "engagement", "tastes" and so on.  A distinction could therefore be made between social network platforms and consumer platforms, despite the trend towards some combination in many of them. On the one hand, the social registry that Facebook and Google tend to build is akin to what states and social sciences did when designing censuses in an attempt to assemble socio-demographic features of individuals and traces left while using apps and interacting with "friends". On the other hand, consumer platforms are more dedicated to a sort of mapping of affiliations, types, styles, and preferences, all of which are more ephemeral or more precisely cyclic than the traditional social features. This is quite close to the traditional social study of moods, fashion, and opinions in political life or in any marketing area, which consist of individual expressions, aggregated and mapped to make the opinion live, or of  trends in cultural tastes. However, it is true that the social sciences and marketing research methods did use sampling and were indeed concerned about maintaining the connection with the socio-demographic features of the individuals they selected in their samples, which is not true for Amazon or Apple and any other consumer platforms. These two sets of digital data (socio-demographic features on the one hand, with Google + and Facebook, and preferences and tastes on the other, with Amazon and Apple) are not so new, indeed. Due to the specific way of producing them in a natively digital format and sometimes on the fly, they cannot be considered as substitutes for other protocols such as censuses or surveys. This is nevertheless how they come to be used by many data scientists and by some social scientists seeking shortcuts to the social. Twitter plays it own part, which paves the way for social sciences of a third kind, accounting for the high frequency propagation processes.

---
[2] Dominique Cardon has proposed a typology consisting of links, clicks, likes and traces (Cardon, 2013)

## 2. The making of 'society'

As mentioned above, computer scientists do not hesitate to make use of social science categories such as communities, networks, and so on, even though they very often lack adequate knowledge of these concepts. This is a part of what I call "algorithmic positivism". Yet one cannot blame them for the use of a concept like "society" since there is nothing more "taken-for-granted" among members of society. This is why it may be risky to adopt a constructivist viewpoint and try to convince data scientists that "society" was designed and promoted in such a successful move that no one dares to question it. If we want to understand the historical times that we are living in as regards quantification methods (and the social reflexivity that is attached to it), we need to look back to the times of the construction of this entity, "society". Let us pretend here that Durkheim succeeded in making "Society" exist. The term was not coined by Durkheim, obviously, although its history is not a long one. The archaeology of the concept of society (Latour, 2005) could be further enriched by calling upon the work of Quetelet (1846), who produced the "average man" which long remained the key to all statistics. At the end of the nineteenth century, however, and largely thanks to Durkheim's genius, "society" took a strong stance regarding "community", which was still prevalent (see Tönnies, 1887). Durkheim's early work on the "division of labour in society" (1893) was not based on statistical methods, but instead laid the foundation for a model of social types, aggregated in mechanical and organic solidarity. Detailed examination of legal systems served as demonstrations and therefore relied on the groups formed or being formed that are legal systems in their traditional or more modern aspects. With "The Suicide" (1897), the method was set up to extend the discussion of the types that would reveal anomie to be a problematic situation. But reliance on data records produced by states, from their various components (ministries, prefectures, governments) became key to the demonstration. It was these aggregates that are explained or explanatory, using a method of comparison between countries, regions, counties or districts, where possible and necessary. The method depended entirely on the available data and could not afford to criticize or to question the procedures for the production of this data, despite the countless limitations identified upon publication. By organizing all his systems of proof around these national administrative statistics, Durkheim found a quantitative analogue for his conceptual choice that put "Society" in a separate status from all manifestations and individual behaviours. Durkheim's *whole* became an entity of the second degree, "Society", (Latour, 2005), while the censuses and other state-data-registers simply perform the task of recovering individual, administrative events (marital status, judicial procedures, etc.), formatted in identical categories and aggregated to reveal the behaviour of populations. Durkheim's strength of conviction would have been to make these statistics exist as equivalent to his "society", where the quantification is able to account for a "whole" through the quality of exhaustiveness, while the concept accounts for the agency of the social structure as such.

It is necessary to note that a form of "convention" was formed between data producers from the state administrations and the emerging social sciences. Together they produced the entity "society" as the object to be tracked by the state for the purpose of governing and to be explained for scientific reasons. The result is the widely shared and obvious fact that 'society' exists, and the methods that allow it to do so have no grounds to be questioned because they demonstrate both their scientific and their operational value: they are "tools of proof" and "tools of government" as Desrosières (2014) put it.

## *The age of calculations and calculators*

Other historical proximities are noteworthy, that do not mean causality but that do allow for an understanding of the power-gains this approach affords in making society exist. In 1890, Hollerith used his machine (that he had invented a few years earlier and for which he filed a patent application in 1886) to conduct the U.S. census. The Census Bureau had not yet finished processing the previous census dating back to 1880 when it had had to start the next one. A change in technique was both necessary and available. Hollerith's tabulating calculation-machine did the work and was sold for doing censuses in several countries. His company would later be transformed into IBM by Watson, in 1926. We can see how the power gained in the counting and description of populations reinforces the status of the State and offers it supposedly useful sources of information for its governance. The pretence of the calculation's exhaustiveness seemed to fulfil the promise of the concept of society: a technical device capable of inputting all that existed, that as Hollerith's census-procedure-equipping machine. It should be noted to what degree the *investments of form* (Thévenot, 1986) that are censuses, end up being events in a given population and become indisputable, appearing almost ritually, to portray all that is social onto the members of this society

Alain Desrosières had amply demonstrated this process by showing how Durkheim's concept of society took the genesis of nation states into account. Nation states rely as much on these figures as they do on infrastructures. From this point on, territory became a key mediation, of which we find traces in the emergence of social welfare policies or in the developing national commodity markets, owing to the railroads, and then in national electoral campaigns through the media (the press also circulated via the railroad, after which came radio, that would become a vector for the emergence of opinion).

Operations conducted by Durkheim have had an impact on the production of other 'wholes' because he constantly insisted, in particular against Tarde, on the separation of sociology from psychology. The latter now works with another 'whole', the individual, which can be measured by cognitive psychology with the aid of experimental devices. The supposedly real 'economy' was entirely performed by economics (as shown by economic sociology, Callon, 1998), thus creating a fiction separating it from the web of relationships in which it needed to be re-embedded. Value has become the key to measurement while currency is actually the powerful medium (Orlean, 2011). Anthropology tried to make the *whole* exist with "culture" too, but resigned itself immediately to cultures in the plural, reproducing the distinctions between French and German traditions, which did not allow for the striking force of a concept yet was certainly useful to the colonial enterprise. Political science adopted this *whole* for the "state" as a reference, and the difficulty in admitting the expansion of politics outside of institutional spheres indicates that this concept has kept its definitional power for the discipline itself. But in modern Western countries, the State gained its legitimacy through electoral processes that rely on nations, those "imagined communities" (Anderson, 1991) that work as content while the State is the container (Boullier, 2011), or the "materiality" and the "statement" that constitute the apparatus in Foucaldian terms. This is why "society" is always enacted in various "nations" although social theory tries to extract it from the limitations of national boundaries. The first generation of social sciences was indeed doomed to methodological nationalism (Beck, Sassen) and still has problems inventing the methods to account for a globalized world made of flows (finance, media, commodities, migrants, and so on). However, we shall see that when examining this digital world, it is quite difficult for us to escape "methodological platformism", due to the total dependency on the traces data platforms deliver!

Durkheim's achievement has been to form an assemblage of very powerful mediations:
- Censuses
- Assembled and formatted by public administrations
- Under guarantee of exhaustiveness

- For States
- For government purposes
- To produce "society"
- Using tabulating calculation machines

## 3. The construction of 'opinion'[3]

The contemporary situation is undoubtedly not that far from another key moment in the history of the social sciences that would help us to understand both what is happening and the conditions of felicity of new conventions. This is why we will consider some features of that period when opinion polls were invented, as an indication of the equivalents in our times. After the first generation of quantification, used by Durkheim to build the concept of "society", we could give the label "2G" to the emergence of public opinion in the late 1930s. In 1936 George Gallup was able to predict the election of Roosevelt over Landon, based on a survey of 50,000 people. Roper and Crossley had done likewise at the same time. Gallup not only impressed the media and policy makers, he radically disqualified older methods (straw polls), including that of the Literary Digest, based on responses from 2 million people, whilst even predicting their own erroneous results (Osborne and Rose, 1999). This impressive demonstration lays the foundations of the survey's reliability and of investigative sampling methods. The exhaustiveness of inquiries on entire populations was indeed sacrificed in the process, but the new approach managed to produce accurate results, provided that the terms of *representativeness* were respected. It nevertheless failed to predict the victory of Truman in 1948, whose voters changed their minds in the last ten days. Methods thus applied to political life and to life-size tests as important as a presidential election had previously been tested on readership studies for which Gallup had operationalized stratified sampling. In fact, these methods had already been applied by the Norwegian Kiaer in 1894. Similar statistical methods in the field of agriculture and later in unemployment in the early 1930s, in the USA, underwent profound changes, from the correspondents' method to random sampling based on probabilistic approaches (Desrosières, 2001) – as Emmanuel Didier has also shown (Didier, 2009). Quota methods based on "sensible choices", where the selected sample as matched with certain properties of the population identified by the census, were however different from those methods of stratified random selection, and were even despised somewhat by statisticians[4] (cf. Stephan quoted by E. Didier). The data collected were also very different, since statisticians from agricultural or employment administrations wanted to obtain "facts", but were nevertheless obliged to rely on statements, not measuring machines, even if they attempted the latter with the "crop meters." Yet the sampling-legitimisation-operation generally succeeded, primarily thanks to Gallup's performances (1939), which were dedicated entirely to other social worlds, those of "public opinion" and not "society". The latter remained a reference of statisticians of the federal state and its offices. It was unquestionably in the context of the mass media that the importance of sampling was recognized. With Ogilvy, Gallup studied film audiences, and then with Crossley, at Young's and Rubicam's, he studied radio audiences, using telephone interviews before even making a proposal to conduct the election polls. From this point of view, Gallup's name must be associated with that of Lazarsfeld, who in the same period, in 1936, launched a "Radio Research Program", based on audience-research-work begun in 1930[5]. Together with Merton they launched the *focus groups* method as early as 1941, and their study of Decatur in 1945 provided the data for the analysis of "Personal Influence" published in 1955 (Katz and Lazarsfeld, 1955). The latter study established the framework for analysis of the "two-step flow" in which mass media play a role, but through the mediation of various kinds of influential relationships.

---

[3] The works of Osborne and Rose (1999), Loïc Blondiaux (1998) and Joelle Zask (2000) develop this story extensively.

The links between the mass media and politics are thus elements of new statistical methods. As Alain Desrosières noted (op.cit.), a national election's predictability actually depended upon the formation of a common public media-space across the United States, and only the radio could do this in such a way as to make the state of voter knowledge about electoral candidates comparable. Considerable media transformation and the mass media (radio at the time) established the conditions for the emergence and validation of a survey technique, which thus opens up a whole new era, most notably for political science and market research. Moreover, it is "public opinion" itself which takes on a measurable existence with these sampling methods whose performative power will by far exceed their experimental phase.

## *Markets and national publics: the scale of the media*

The missing link in my description remains the vehicle of financial incentives for such investments, needed to understand a public. Communication agencies such as polling organizations cannot live solely from their campaign activities, even if they do bring them high visibility and renown. From the outset, their target was the mass media, as noted above, for one essential reason: audience measurement has been the key to the distribution of advertising space, since the dawn of radio and then later with television (in 1941 the first advertisements were aired on American television for Bulova watches, during a baseball game). But these measures also serve to monitor the impacts of these campaigns on the minds of consumers, giving an unprecedented boost to marketing, which in turn drives increasingly sophisticated communication strategies (Cochoy, 1999). Brands have thus been present, from the outset, in methods of inquiry into opinion using sampling; that is, from the moment such investigations were aimed primarily at mass-media audiences. Market research on consumer goods developed at the same time, from the 1930s, and in the same movement of national standardization of products, as Desrosières pointed out. The production of a unified national territory, through the media, that included transportation and mail, established a new condition of felicity for these survey methods. This allows me to draw a direct parallel with the recent creation of a global market, this time, through the domination of digital platforms. Google, Apple, Facebook and Amazon have produced the same effect on a global territorial scale as radio and the railway had on the territory of national markets. This is in line with the work of McLuhan (1964), for whom the change of scale in itself constitutes another world far beyond the property or goods exchanged.

---

[5] The coupling of operational / academic consisted rather of Gallup-Cantrill on one side and Roper-Lazarsfeld on the other, but history has remembered mainly Gallup and Lazarfeld. See Blondiaux on this topic.

## *Public opinion exists, I measured it!*

The work done by Gallup on the operational side and Lazarsfeld on the scientific side is therefore not a simply a marketing operation or a face lift for the social sciences: it provides whole societies with methods with which to analyse themselves and to represent themselves – as opinions. Tarde (1901) has certainly highlighted the importance of these views, yet it is only when the metrics are established and produced in a conventional way that opinion finally exists. Only the media's control and their ability to produce a unified public in a national territory enabled this methodological assembly to hold. The "whole" referred to by the polls is in fact originally the *public* formed by the media, which allow the *audience* to emerge as *public opinion* and to make it permanently visible and measurable. This connection between audience measurement and monitoring methods for public opinion, a connection that is both technical and historical, must be regarded as the key to the device: the media want, above all, to measure audiences, as did Gallup for reading, but the techniques in place turned into predictive voting tools, which justified this betting on public opinion. The whole "audience" or even "public" has mutated into "public opinion" and managed to detach itself from its own reference within the media (which measure themselves), for the purpose of being exploitable for brands to measure the influence of their campaigns. The parts (Latour et al. 2012) that are individual expressions are preformatted to be recordable and calculable, but the link between the parts and everything else is made only by the pollsters' black boxes. The rigorous, scientific precautions are upheld through 'confidence intervals' (defined by Neyman in 1934), which keep a reference on the comprehensiveness of the studied population. Bowley (1906) proposed these principles in 1906, when speaking of the 'probable error', allowing for the clear linking of the polls and the emergence of statisticians' probabilities, as E. Didier does. But soon these precautions will disappear from the findings, as seen in the contemporary media. At this point, everyone knows that "opinion exists", whatever the report about the artefacts needed to make it exist, and despite what Bourdieu said about it (1984)[6]. It has been naturalized, "taken for granted", and the sampling methods lie buried beneath the powerful performative effect of these immediate, aggregated indices. The approximation remains acceptable, especially with the repetition of the same questionnaire over time (by panel, independent rotating sample) under identical conditions, "all things being equal". It allows for the smoothing-out of biases which then become acceptable by convention. Such successful convention work focuses on the same assemblages of mediations already mentioned for society:
- the "surveys" and "polls" (from individual expressions framed by questions and thus made calculable)
- assembled and formatted by pollsters
- guaranteeing the representativeness of samples (sampling)
- for the media
- for the purpose of monitoring
- to generate public opinion (and audiences).

---

[6] Pp. 222-235. Stating his thoughts at the end of the article, he writes: "Public opinion in this sense, implicitly admitted by those who carry out opinion polls or those who use the results, I'm just saying that *this* opinion does not exist". Being the champion of social structure and its reproduction in long-term trends, Bourdieu could not allow room for other entities such as public opinion, which is more cyclical, to exist independently.

As Alain Desrosières (2008) put it, the essential thing is not whether these data are a reflection or mirror of society or something else, but rather whether they "make something that stands by itself". Note that there is a new element at work in this chain: that of the methodological limitation, expressed in terms of the representativeness of the samples, because this element is still missing in digital traces, which explains much of the uncertainty and suspicion on all results compared with the polls, for which the "biases" are well known but have been controlled by convention since the 1940s. The "consolidation" that Emmanuel Didier (op. cit.) describes for statistics and surveys remains to be done. This rather long account of the successful fabrication of opinion was necessary not only to understand the similarities between that "epoch" and our current one, but also to measure the distance and the work required to produce conventions of equal quality (Eymard-Duvernay, op. cit.) that would make "traces" exist as entities recognized by the social sciences. We certainly need to consider opinion to be a social reality that lives its life and is no longer called into question, thanks to the quality of technical and institutional arrangements that stabilized its mode of appearance, notwithstanding the many critiques that they still face. To be sure, the worlds of social science and marketing differ, yet for years they have used the same methods and even the same samples while being able to distinguish themselves from one another. The question posed by this new world of traces emerging on the web is of the same sort: how can we invent the social sciences in a way that suits these traces whilst admitting the conditions of their production and utilisation?

## 4. Three generations, three points of view on the social

Building on these two historical landmarks of society and opinion and on the mediations which managed to make them exist as taken-for-granted entities, I propose a comparative table in which the next generation of social sciences is designed along the same criteria in order to devise some roadmap for the consolidation of the conventions. Digital devices generate new opportunities for quantification but not only in terms of volume. My main statement is to emphasize the emergence of new entities, apart from society and opinion; entities that have an agency of their own and that I call "replications", propagated along networks in the material aspect of traces.

The following table will make the comparison more visible and will introduce my understanding of the stakes of third-generation social sciences.

**Table 1: The three generations of social sciences**

| | 1st generation | 2nd generation | 3rd generation |
|---|---|---|---|
| **Concept of the social** | Society/(ies) | Opinion(s) | Replication(s) |
| *Collection devices* | Censuses | Surveys/Polls | Platforms |
| *Validation principle* | Exhaustiveness | Representativeness | Traceability |
| *Co-construction institutions/ research* | Registers/ inquiries | Audience/ Polls | Traces/ Repurposed digital methods |
| *Major players of reference (and funding)* | States/ Nations | Mass media/ Audiences | Platforms/Brands |
| *Operational actors* | National Institutes | Polling organisations | Web platforms (GAFA) |
| *Founding authors* | Durkheim | Gallup, Lazarsfeld | Callon, Latour, Law |
| *Key problems of scientific approaches* | Division of labour and the welfare state | Propaganda and media-influence (audience measurement) | Science and technology (scientometrics) |
| *Technical conditions* | Hollerith's machine (tabulating calculation) | Radio and telephone | Internet, Web and Big Data |
| *Semiotic formats* | Crosstabs and topographic maps | Curve and bar charts / pie charts and topology of influences | Graphs, timelines, dashboards |
| *Metrics* | Statistics | Sampling | TPS (tweets per second) (scores) and similarity matrix |
| *Technical criteria for data quality* | Relevance, accuracy, timeliness, accessibility, comparability, coherence | Confidence intervals Probabilities | Volume, Variety and Velocity |
| *The social science's dominant modalities* | Explanations | Descriptive and predictive correlations | Predictive correlations, memetics |

The excessive coherence of any table must not make us forget that what is at stake is the construction of a proposal for the third-generation of social sciences, which is in no way guaranteed. Actor Network Theory has indeed laid the foundations, building on the methods of scientometrics (and previously of bibliometrics), and Tarde did announce the principles as a general theory of imitation (including opposition and invention, not to be forgotten). But, for now, the trend is more at the "end of theory" and the occupation of the field by the Web platforms (GAFA). They produce, calculate and publish themselves on these traces, and all for commercial purposes primarily because major brands are demanding these approaches. I shall describe the choices available to the social sciences in such a context. As the table suggests, not only does Big Data challenge the status of social sciences in terms of empirical capacity and of modelling without theory, but Big Data really needs Big Theory for the social sciences to keep their role alive. This means that the Social Sciences have to experience the rising world of Big Data as anthropologists would have done with new continents – anthropologists who would adopt a more "perspectivist" approach (Viveiros de Castro, 2014) than the traditional modern (Western!) anthropology. The social sciences are indeed "affected" by this new "continent" in terms of both sources and principles of validation, but they gain something in the deal, i.e. the access to a whole new dimension of the social, which used to exist before, but without any tools to trace it: replications, vibrations and the "high frequency" speed of social life.

What begins as a historical comparison, where the succession of generations can let us believe that the next one makes the previous one obsolete, is turning into a more diplomatic statement, where each approach is able to grasp one specific aspect of the social that others cannot account for. This derives from the social studies of science led by Bruno Latour, who has deeply transformed our understanding of the process of scientific knowledge. The insistence on the agency of devices, of scientific devices, can be used for social science as well, because one could say "we've got the sciences of our devices". The census and the polls have built entities that highlight some specific dimensions of the social which came to be framed as "society" and "opinion", but digital devices make something new appear.

### *Generations, waves or viewpoints on the social?*

From this historical account and from this diplomatic move, a general pattern emerges. The social sciences adopt a rather limited number of perspectives related to these devices and to the entities that have been constructed. In short, "society" is generated by a "structure" approach; "opinion" is produced by a "market" approach; and "replications" (those of the digital world) are discovered through an "emergence" approach. When trying to account for the various controversies in the social sciences, one will eventually revert to this classification. And, more importantly, these old disputes can be considered as "viewpoints" on the world, at the epistemic level, equipped with different devices and targeting different entities. We may talk of these differences in terms of "wave lengths" as we did previously: "structure" analysis focuses on long wavelengths, "market" on mid-term and cyclical ones, and "emergence" on high-frequency ones. This should remind historians, with a slight translation, of the Fernand Braudel's famous distinction: long-term history, cycles, events (Braudel 1996). He was right not to condemn any of these viewpoints for the sake of another, and only to advocate sufficient diversity (against the trend, at the time, of focusing on events). All three these viewpoints could easily fall under a general theory of attraction, as accounting for the social in general: long-term attractions of social traditions, habits, and repetitions (the ones sociologists love to love); cyclic attraction of fashions, political opinions, tastes and preferences in general (the ones economics, marketing, psychology and political science are fond of); and high-frequency attraction of replications that make the buzz as well as the financial speculation. Imitation is not a matter of long-term structure, because it emerges at the very moment it occurs, nor a matter of strategic decision, because the time lapse is so short that reactions are quasi "unknown" to the subject, for these subjects are only the targets of replicators, entities that connect every mind in a millisecond. This existed before the digital continent emerged, such as in "olas", crowd moves, rumours, and so on. But no social science has ever been equipped for documenting these contagious processes (Sperber, 1996) in which speed is so critical. Today the platforms amplify these processes and produce the leverage for measuring and tracking them.

### *The wawelengths of JeSuisCharlie*

Let's illustrate this with a discussion about "Je suis Charlie", the global movement that followed the Charlie Hedbo and Hypercasher attacks in January 2015 in Paris. A controversy emerged around the book of historian and anthropologist (expert in demographics) Emmanuel Todd, a rather famous and maverick kind of scholar. He rushed to publish his book (*Qui est Charlie?*, in May 2015) based on the correlation between the number of protesters in the streets in France and the age-old history of Catholicism in each city. The book used well-known maps of the conflictual period of the French Revolution that were based on priests' acceptance or rejection of the new regime. I shall not get into the discussion of the main hypothesis, that demonstrations were stronger in the regions where Catholicism used to be strong and are a significant anti-Muslim signal of this "zombie Catholicism" in France, where traces remain although the active practice of the religion has fallen sharply. It might be true and many previous books by Emmanuel Todd were convincing enough to warrant trust in him once again. However, the method is somewhat unbalanced in favour of long-term waves: the only indicator used for the analysis of the demonstrations is the number of participants estimated both by the police forces and by the leftist newspaper *Libération*, in a sort of reciprocal compensation. The figures are "botched" as the author said, but will do in this state of emergency (in order to deflate the mood of unanimous republican celebration). What is more interesting, is that he did not consider it to be of any interest to ask the demonstrators any questions since, as he said, "very often, they did not know how to explain their participation", as all of them were "carried away by the mimetic intoxication of a saturated media space" (p. 21). The three generations in the same sentence, it seems: surveys do not make sense with "cultural idiots" such as demonstrators and their "opinion" does not exist since the media are reaching a level of excitement that is contagious (mimetic intoxication), and this does not need any investigation, except a critical stance from the point of view of the true and only social science, the one based on long-term indicators that crush all the mediations of events and of opinions in a single move. Of course, French experts in political science[7] clearly documented the capacity of sampling surveys to account for a part of their motivations and their background (surveys were conducted on a large scale right after the demonstration). And Twitter specialists published dynamic maps of the contamination of the hashtag "#JeSuisCharlie" which managed to propagate all over the world in the following six hours. This global dimension of the process was not relevant for a social scientist who relied

---

[7] Mayer, Nonna and Vincent Tiberj, *Le Monde*, 19 May 2015, Rouban Luc, Note du Cevipof, Paris, 13 May 2015

on its "methodological nationalism", due to the idiosyncrasy of a general cause such as the French Revolution. More significant, is the fact that even at the level of the demonstration, the specificity of the event was missed by the author: when two demonstrations took place, as in the case of Marseille (one for the left, one for the right), the score was just an addition of the two, as if a common demonstration does not make a difference in such circumstances. The limits of the model when applied to Paris or to Strasbourg were acknowledged, but never could they question the model itself. The demonstrations did not have a life of their own; they were unexpected emergences of a "collective" that would be quite difficult to label ("the people"?), and for sure doomed to disappear the next day, but were still a striking experience for the participants. These demonstrators were just a number that can be correlated to the only deep, real and permanent causes of French political behavior, two centuries later (with this delay explaining the "zombie" rhetorical trick, since the mediations – religious behavior – disappeared from the long-term radar!). For opinion experts as well, the use of polls was the only way to account for this sudden change in the mind of so many people, and from their individual expressions they were able to build an image of what "public opinion" was saying – a much more complex view than the simple causalities of long-term social sciences. However, they did not make any use of the tremendous amount of data generated on internet, via Twitter, Facebook, Instagram and other social networks, and on blogs, media websites' comments and so on. The public's emotional experience was expressed not only during demonstrations but also earlier, immediately after the attacks, through a contagious extension of some hashtags and especially through the propagation of the meme "**#JeSuisCharlie**". Thanks to the traceability provided by social network platforms, the global phenomenon of contagion becomes visible in real time, as shown on the following map and timeline.

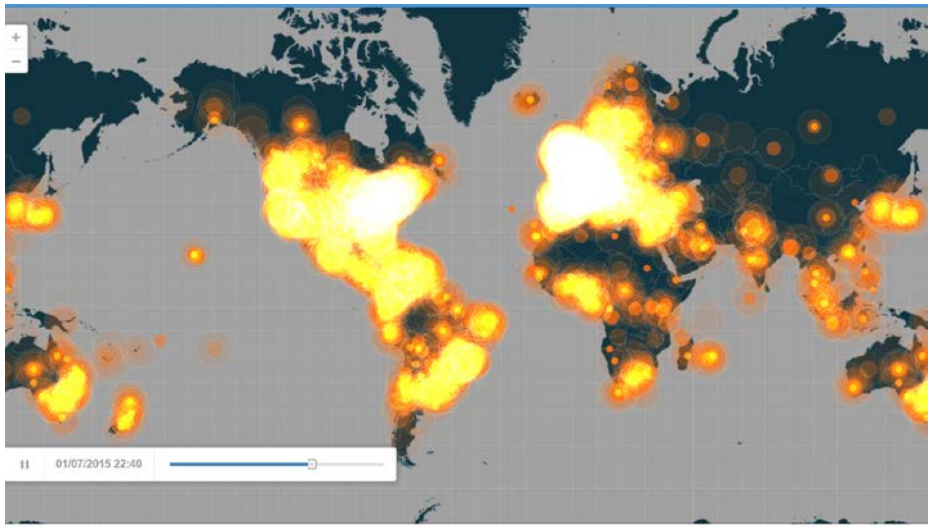

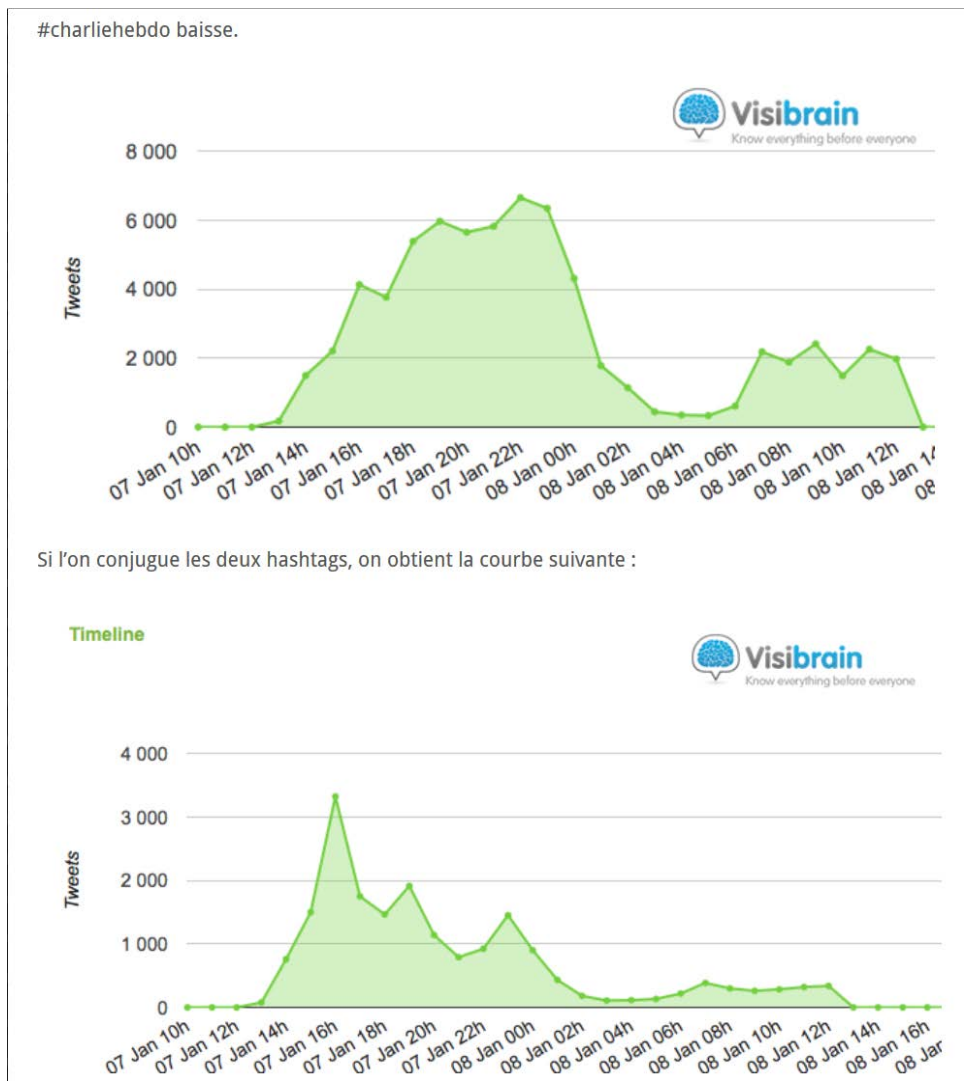

These traces data seem to provide very little information except for the volumes that they reflect: nothing about the Twitter accounts, profiles or networks; just a very simple display along timelines, showing how strong yet ephemeral these kinds of propagations are, and another very basic spatial representation, powerful for demonstrating the extension of the contagion worldwide but not precise enough to allow correlations with any other spatially referenced dataset. This is enough however to trigger more investigation without discarding such a global phenomenon, and to challenge social sciences to account for these high-frequency and short-term waves without using their traditional tools and concepts. All three points of view can bring some insights, none of which can be claimed to be the ultimate cause – in Tarde's critical terms, the *"cause-finaliers"* –, or to assemble all of them in an overarching vision of a "whole" which is just an effect of the devices we adopt to delimit the social life continuum (Latour and al., 2012). Some may look for the translations from high-frequency traces (expressions or hashtags) to the cycles of opinion or to the long-term memory of an event that will become a part of the globalized collective experience. Why not? But there is so much to do first to understand the specificity of these waves of traces and to be sure of what can be extracted from them, that a precautionary principle should apply to the reconnection of all these wavelengths. In order to do so, we need to build the conventions that will guarantee "sufficient scientificity" to the social sciences of the Third Generation.

# 5. Brands' grip on traces

This may look like a gamble since all these traces are produced and controlled by digital platforms and, ultimately, by the brands that are funding these social networks for their own purposes: is there a way to escape the "methodological platformism" (or "platform bias", Marres and Weltevrede, 2013) that can be observed for instance through computer and social scientists' focus on Twitter? Where does this fascination with traces – as opposed to data from registries and surveys – come from, despite their limitations? The traces are actually a key resource for brands to monitor the impact of their actions on the public. Reputation and renown no longer translate uniquely into audience measurements; this would be a simplistic import of measures built for the mass media. On networks, one must measure not only a form of audience (the reach) and the most basic activities of its uncertain public (likes, stars), but also more sophisticated activities such as comments, which constitute what is called "the engagement rate". Brands are fond of these traces and it is they who fuel the turnover of all these platforms, and thereby of the entire Web. The opinion mining and sentiment analysis tools (Boullier and Lohard, 2012) are thus the answer to "the marketer's anxiety after the product launch". However, the extension of this brand domain reaches into all activities, whether commercial, cultural, political, institutional or even interpersonal when veryone must measure her excellence with rankings, as researchers are requested to do (Bruno and Didier, 2013). It is the brands' methods that take precedence everywhere and impose their law and their pace, even on public services. But what concerns these brands primarily is not structured and constructed data to test e.g. causality, but many traces that function as indicators and alerts, even approximate ones, not at the individual level but at the level of trends. Similarly, it is not reflexivity that is sought but primarily reactivity, the ability to determine which lever to act upon in relation to the dimensions (features) of the brand that are affected. The political world itself is now caught up in the spiral of reactivity and its addiction to tweets led us to consider that we have entered the era of *High-Frequency Politics* (Boullier, 2013) in the image of the *High-Frequency Trading* of speculative finance.

Platforms pick up traces of the actions and clicks of Internet-users or machines, in a standardized format that aggregates them and produces a score. This score is displayed and can be used by the platform itself to show trends to guide the placements of advertisers who also seek to achieve certain effects and to optimize their investment or communication choices. In a simplified format, this is the string of events that was produced. The performative mechanism works almost identically to the audience measurement. Some then try to develop a critique showing that the "likes" aggregate very different sorts of behaviour, including even purchased likes. The limited quality of the traces is observable on all platforms, but these limits may be intrinsic, when they do not meet the criteria for traceability that we consider crucial in order to exploit them, or extrinsic when we criticise their lack of reliable relation to the "real" world. It is the latter stance that we find in boyd and Crawford as regards Twitter: "Some users have multiple accounts. Some accounts are used by multiple people. Some people never establish an account, and simply access Twitter via the web. Some accounts are 'bots' that produce automated content without involving a person. Furthermore, the notion of an 'active' account is problematic. While some users post content frequently through Twitter, others participate as 'listeners'. Twitter Inc. has revealed that 40 percent of active users sign in just to listen." (boyd and Crawford, 2011). Other studies (Driscoll and Walker, 2014) tested the data produced from various access methods offered by Twitter, for example, and showed that the Search API, the Streaming API and Gnip Power Track (paid service) provide very different results. The latter method for instance collected a much larger number of tweets, but not uniformly according to the requests! This means that the traces collected are entirely dependent on the collection mechanisms , which is not surprising but which we do tend to forget since other, older methods have become conventional.

But these limitations hardly concern operators, platforms or advertisers. Their action / reaction works in the performative mode, where the likes reveal a reality that will initiate strategies to influence the likes, in a self-referential cycle to which one could also assign audience ratings. However, in the case of audience ratings, all advertisers and programmers have agreed on stable criteria and produced a shared agreement, and evidence of this has come to forcefully impose itself every morning in the management of programmes in the mass-media. Social network platforms and advertisers have not yet reached a stable compromise, which explains the proliferation of services that claim to be the standard, as I have shown in the case of Klout (Boullier and Lohard, 2015), and that want to become the Nielsen of these measures. It is easy to see the difference between these principles and the traditions of the social sciences, as G. Bowker does, and to show their extreme reductionism: "If I am defined by my clicks and purchases and so forth, I get represented largely as a person with no qualities other than 'consumer with tastes'. However, creating a system that locks me into my tastes reduces me significantly. Individuals are not stable categories – things and people are not identical to themselves over time. (This is argued in formal logic in the discipline of mereology and in psychiatry by, say, ethnopsychiatry.) The unexamined term the "individual" is what structures the database and significantly excludes temporality." (Bowker, art. cit.)

Bowker has cause for concern from the point of view of "society", but the third generation of social sciences is not so much interested in "society" as in other social processes created by other devices, but which, nonetheless, cause us to act. Brands, reputations and recommendations as they are exploited by Amazon can certainly be forcefully re-injected into a matrix "society" to make them say what they are not made to say. But they also say something of themselves, from another world, that of the power of recommendations and contagions that the social sciences are reluctant to understand. It is as if the sociology of "society" were reliving an analogous experience to the one that anthropology provoked, that of the necessary shift with the modern world and its categories. To be sure, Durkheim did first use it to analyse religions (Durkheim, 1912) and to employ traditional societies and totemism for his demonstration of the power of society on individuals. But Mauss (1950) made a great side step in recognizing the power of things and the spirit that persists within them, the 'mana', for which Levi-Strauss criticized him. Another side step was taken by Descola (2013), demanding that the sciences in general be put into one specific (naturalist) ontology among several available worldwide. With Viveiros de Castro (2014), the challenge is pushed further still, since his anthropology adopts a perspectivism that includes its own foundations, once it agrees to be affected by the world it observes, namely collectives establishing "assemblages" that are far removed from Durkheim's principles of "society".

Let us take the risk of behaving in the same way that these anthropologists have with the world of big data, with the Internet of Things, with correlations which spawn multiple links that science would wish to purify and reduce to causes. The world being born is as different from the modern world and the Social Sciences as it was for Descola's Achuar people, if one accepts, as he did, not to project all our categories onto something but to think from within and to 'become with' them (Haraway, 2003). So it is useless to complain about the imperfection of the data and its approximation because we are now dealing with traces through a process of pervasive traceability. This traceability connects entities that did not exist beforehand, but are now endowed with an IP address (thanks to the availability of IPv6, or $3.4 \times 10^{38}$ addresses) and so can interact just as humans do via their machines. The vital statistics which are the reference base for the 3rd generation of the social sciences are no longer the censuses but an index of IP addresses, totally agnostic about the entities that "are behind" because all act almost equivalently and cause the others to act. This shift may seem radical, but it helps to hold together the approaches of previous generations while watching the ever-present world with the tools and relevant categories at hand.

I have drawn up a table that merits systemisation. Digital networking generates:
- traces
- assembled and formatted by platforms
- for brands
- with a view to reactivity
- in order to produce rankings or patterns.

This situation is akin to the two other key moments in the existence of the social sciences, especially sociology and political science, discussed above However, these new methods and principles have still to be arranged in such a way that they transform themselves into "socio-technical conventions".

# 6. What the social sciences can do with digital traces

Setting digital transformations in the long history of the social sciences allows us to better understand contemporary movements in the use of traces. All three approaches or wavelengths can still be applied when addressing the dataset provided by digital platforms. The choice between structure/market/emergence is still significantly present in the digital methods adopted by those social scientists who are the most eager to dig into this ocean of data. I shall list the three dimensions of the use of traces data and focus mostly on examples from Twitter studies. By doing so, my intention is to follow the insights of McKenzie and Vurdubakis (2011) for a more cultural understanding of code as a "will to power" and a "will to knowledge".

## *6.1. "Society" and digital traces*

The first direction consists in taking up the well-known methods and concepts and applying them to traces collected on the web or, even more sensibly, exploiting the potential of digital networks to implement exactly the same methods. Censuses could be equipped with computer terminals to speed up and standardize the collection of data, which makes the now visible 'society' even more reliable. Public registries that track all activities for administration purposes can build up databases and apply machine learning methods to match crime reports and building vacancies as did the New York City office. This trend is amplified by the pressure for public open data. However, the concepts, the models and the entities can remain exactly the same as those used in pre-Big Data settings, except for the use of machine learning, which is not yet very well documented. 'Society', as a statistical and theoretical entity, is gaining more solid foundations as data series become more long-term, precise and interoperable. But even Web studies, a by-product of the social sciences, can also implement the same framework on these new media formats. Think of the extended use of Google trends for any topic, of Google queries for matching behaviours and well-known regional differences, of Google n-grams for the emergence of vocabulary, or of Google books for long-term study of literature. The works of Francesco Moretti are the most famous in this field because he coined a new concept and methodology, "Distant Reading" (Moretti, 2011), to address new issues in the history of literature. In order to do so, he had to quit the focus on texts, style, and established trends, and to take the huge numbers of unknown (although published) novels at their face value, that is, their titles and authors, dates and places. What he called "the slaughterhouse of literature" (in a paper published in 2000) has been revealed and can become of some value provided that a kind of "surface analysis" is accepted, by extracting "a form" from thousands of novels that have fallen into oblivion: "form is the repeatable element of literature", "devices and genres, two formal units". The radical choice is to accept the limitations of the data available for the benefit of a minimalist but quantitatively sounded approach, to make use of data-mining algorithms to handle this volume and to learn from emerging patterns. By doing so Moretti is, in a part of his work, keeping the same commitment towards long-term history but at the same time agreeing to question all categories so that some unknown patterns and issues can emerge. This is because he remains very careful about not being trapped in any "algorithmic positivism", as I call it. To emphasize this stance, he recently advocated some "hermeneutics of noise", the kind of oxymoron that disturbs computer scientists.

When computer scientists work on Twitter datasets or Wikipedia ones, for instance, they really value these kinds of sources because they have boundaries, they keep track of everything and the "whole" can be designed as a graph, irrespective of the entities that are used as nodes for the start (accounts, tweets, pages, and so on). The preference for Twitter and Wikipedia in these studies is also due to the relative openness of their API, although Twitter is limiting access by offering Gnip API and making all users pay for this access. Here is one limitation to keep in mind: social sciences cannot do without these commercial arrangements with platforms; they must accept all the social models that are encapsulated within the algorithms of the traces data collection. But before the digital era these arrangements had to be negotiated by social scientists with States and their national statistics bureaus, or with pollsters. From "methodological nationalism" to "methodological platformism", not to mention "methodological mediatism", for opinion makers, there is no way to detach from the material conditions of production of knowledge, except by being aware of them. As far back as 2008, one of most talented and pioneering teams working on social networks, from KAIST (Ahn and al. 2007), studied Cyworld, the well-known platform in Korea. They focused on the "structural characteristics of the activity network compared with the friends network" and demonstrated that "its structure is similar to the friend relationship network". The dynamic part of the activity was thus correlated to the social structure of friends that is so easily computable since SNS produce nodes that are 'friends' and allow graph calculations. This is a general feature of SNS analysis, the use of graph analysis to find solid ground representing a "whole", and is a useful tool for displaying social positions. Most of the time, the reference to social network analysis is only superficial but there is clearly a connection that could benefit this rather recent tradition in the social sciences. In fact, the term "network", when dissociated from the concept, is rather misleading. Network analysts' networks are actually often complete networks and one understands the imperative need for exhaustiveness. For diffusion networks (Rogers, 1963) the range of configurations studied is limited, partly for practical reasons, but it does allow one to claim generalization through a form of representation, because comparison allows for the eventual correction of any of the distorting effects of an overly reduced sample. Moreover, the concepts used in these approaches must primarily be used to further questions in terms of power, inequality, issues of a structural nature for the "whole" that is society. This makes it easy to understand why network analysis has barely begun to exploit the Web as it exploited these concepts in the context of the first generation social sciences. The same phenomenon managed to be partially repeated with scientometrics, for which network analysis also represents a resource but has developed specific approaches alongside it. Scientometrics (Callon et al., 1986) has the huge advantage of working on a body of data that has been normalized by a community which has done considerable disciplined work on itself, through its 'disciplines' specifically, connecting all references to authors, articles, journals, institutions, and keywords themselves. This tendency has been further accentuated with the contemporary trend in international rankings that have become essential both for careers and for the financing of institutions and their attractiveness. The methods used to study citation networks, as thematic developments for example, greatly benefit computing capabilities for digital technology and new methods for graphs. However, they can in no way be transposed to the study of traces because their essential completeness or even representativeness are easily filled in a world as standardized as that of scientific publications while the web provides no framework to navigate, and no reference to any 'whole'.

More recent studies of Twitter, such as Lin et al. (2014), focus on the social status of users and have a looser approach to social structure. When trying to account for "rising tides" (how collective attention aggregates on some specific events) in a more "emergentist" view of the world, the authors finally get back to the unequal distribution of fame to explain the specific role of some "rising stars" on Twitter. "Elites also appear to guard their status, indicated by their restraint in retweeting others at times when both rookies and typicals increase retweeting behaviour, suggesting a reluctance to 'anoint' others as worthy of attention through retweeting their content". And they refer to some kind of "two-step flow" but just to confirm the social places' structure and not to track any real influence or opinion propagation. When Weng et al. (2010) try to detect influential twitterers, they explain it by homophily, a structural feature based on network reciprocity and not on the specific capacity of individuals. They do not even imagine that some tweets could be more influential than accounts. However, their results limit the influence to exposure to content posted by friends. These are the limitations of these works in which influence can be defined in a way that better suits the model. The findings of Weng are explicitly challenged by Cha and al. (2010) and later by Bakshy (2010, see below). Cha and al. wanted to criticize the "million follower fallacy" (as Avnit did, in 2009) by demonstrating that in-degree (number of followers in the Twitter graph) "reveals little about the influence of a user". When changing the type of metrics adopted (retweets and mentions instead of in-degree), they were able to detect that mainstream news organization (retweets) and celebrities (mentions) performed better, and to suggest marketing strategies targeting these top influential personalities (Watts and Dodds, 2007). Which is exactly the contrary of what was to be demonstrated by Bakshy (see below) with a methodology focused more on individuals and on what experts in Web topology are used to practising, that is, erasing the top level of the scores in the topologies, where Amazon, Acrobat or other central nodes always perform better in every domain, for topological reasons.

Some other studies make use of on-line networks and of computer modelling to build some kind of experimental sociology which uses the resources of the social networks without sticking to their internal metrics. For instance, King and al. (2013) designed an experimental method to check which kinds of messages were censored on Chinese social networks. They detected a difference between "criticism" and "appeal to collective action" within the messages, and the latter were most often deleted by the Great Firewall staff. In order to test this hypothesis, they decided to generate messages of their own, displaying these two clearly distinct features, and they validated their first impression. On a larger scale, Savage and al. (2013) designed what they call a "class calculator", a fast quiz for BBC audiences who became interested in the publication of the Great British Class Survey. This story emphasizes the "recursive loop" that was created and which deeply transforms the status of social sciences and their audience. However, this combination of old and new methods does not question any of their assumptions about social structure and classes as major concepts, and the authors do not contest it. As we can see, the amplification brought by digital methods may enhance the first generation or long-wave style of sociology without addressing the new kinds of traces produced by networks.

## *6.2. Opinion and digital traces*

As mentioned above, the global misunderstanding about "opinion mining" is prevalent and, due to this misunderstanding, computer and social scientists consider the harvesting of expressions from tweets or blogs as mining of "opinion". By doing so, they erase all distinctions in the mode of production of data, and they ignore the advice of Ruppert and al. (2013) about digital devices: "we need to get our hands dirty and explore their affordances: how it is that they collect, store and transmit numerical, textual, aural or visual signals; how they work with respect to standard social science techniques such as sampling and comprehensiveness; and how they relate to social and political institutions". "Opinion" was carefully designed as an artefact that works (although some limitations and scepticism are often highlighted) thanks to the mediation of polls and sampling. Despite this precautionary principle, the use of digital traces as substitutes for opinion are prevalent and are very difficult to challenge since verbatims in blogs and comments for instance look more massive, spontaneous, authentic, computable, real-time, im-mediate, and so on. If we use "opinion" as the entry for all approaches[8] that refer to choices, rational choices, preferences, decision making, moods, cycles and so on, most of the studies of natively digital sources such as social media fall into this category. In the tradition of viral marketing, the decision-making process must be accounted for as a product of trend-setters, influencers or opinion leaders, as in "word of mouth marketing" research for instance. In a 1st-generation style, "influentials" may be considered as structural roles in social networks analysis and their level of integration will explain their adoption behaviour, as for instance in Coleman and al.'s (1966) medical doctors' famous work. But they might as well be considered as individuals with a strategic decision-making agency that will be better analysed by the information flows that strike them in general or only one oftheir social features. This is where the opinion models get back to preferences, information completeness, equilibriums and so on.  One study of influencers on Twitter (Bakshy and al., 2011), for instance, chose to bet on the distribution of influence among all users: "everyone's an influencer". The findings are consistent with many other studies that influence can be detected only in specific areas and cannot be of a general kind, but the authors discuss the respective value of using Retweet or Repost behaviours as indicators. They really try to avoid any limitation to network structure for explaining propagation, and their paper investigates individual influence at large and then the role of content (an orientation which is more related to 3$^{rd}$-generation social sciences). They are not able to find any feature of contents that could account for the propagation patterns but they underplay the role of influentials in advising marketers "rather than attempting to identify exceptional individuals, (…) to instead adopt portfolio style strategies, which target many potential influencers at once and therefore rely only on average performance".

By contrast with this rather precautious style of research, Kempe and al. (2013) do not hesitate to use simulations to test cultural dynamics. Agents are equipped with selection capacities as were the neighbours in Schelling's urban simulation on segregation (1971). Their model demonstrates the optimal social distance in a population for the explanation of influence processes (that tend towards homogeneity) vs. selection ones (that enhance diversity). The issue is relevant for political purposes to test whether the trend towards the "filter bubble" (Pariser, 2012) and the "rich-get-richer" effects are validated on internet. Their highly mathematical model tests the distance where the system reaches equilibrium (and the answer is: systems "in which the active nodes have a pairwise distance of at least 3"). Why are we looking for equilibrium and why do we have to simulate systems which by definition require all other features to be seen as externalities? This is a major trend when trying to model individual behaviours and their collective effects, and the influence of paradigms from economics is quite clear, although it is limited to mainstream formalized economics which tries to perform the economy as a separated world as well.

## *6.3 Replications and digital traces*

The social science traditions are less useful in this area of replications for transferring concepts and methods to a next generation, for the simple reason that the study of the events, imitation, and rhythms of social life was very poorly considered before the deluge of traces data on the high-frequency life that weave the social. Take some examples, such as: the vision of Park who wanted to focus on the "movements of crowds", rather than on substantive crowds as Le Bon used to do (Boullier, 2010); the crucial role played by Tarde, whose imitation made of small differences, conversations and hesitations was crushed in the academic world by Durkheim's all-powerful Society; and some insights from geographers (DeLyser and Sui, 2012) about the need for a rhythmanalysis as a social science of its own. Microsociology and ethnography have a tradition of tracking this type of material (such as rituals) but they miss the opportunity to quantify these phenomena due to methodological limitations (Boullier and Lohard, 2012) that can be overcome with digital real-time and high-frequency traceability.

And this was clearly demonstrated by Leskovec and al. (2009) when they designed their meme-tracker and followed the cascades generated in media (social and mainstream ones) by citations during the Obama 2008 campaign. As their definition of memes was largely open, compared to those fo the founders of memetics (Dawkins, 1976), (Blackmore, 1999), I prefer the distinctions made by Shifman (2014) between virality (propagation of a similar content) and memetics (propagation of contents and its derivations). However, Leskovec and al. managed to build phrase graphs that accounted for the relationships (edges) between different versions of the same citation distributed along a timeline, even though their linguistic features are different. They clearly refer to an imitation model but their theoretical assumptions lead to a "preferential attachment" from sources that select the content to propagate. This looks like a more "market-oriented" model but authors consider the recency of items and their attractiveness as key factors in the spreading of citations. This is exactly the shift of vision that social sciences should look for: from structure or market to emergence, in which items, content, and memes display their own agency. Their insistence on finding a pattern to the news cycle leads to their abandoning of the attractiveness track of research which would have needed a more semiotic approach to the memes features. The same limitation can be found in papers which try to assess the way the competition among memes works (Weng and al., 2012). The principle of a limited attention is a promising insight to start with but, according to the authors, the structure of the social network will account for the propagation "without having to assume exogenous factors such an intrinsic meme appeal, user influence, or external events" or "different intrinsic values among ideas". The reluctance to attribute some agency to cultural items such as memes is motivated by the need to stress the equality "among ideas" in a rather strange republican prerequisite, and strangely enough, meme are deliberately considered as exogenous, although they mention that intrinsic features of viruses are relevant within the "conceptual framework of biological epidemics". Why is it that computer science would accept the role played by viruses as agents while discarding that of cultural items like memes? I would explain this discrepancy in the terms of Philippe Descola's four ontologies model: naturalists (we, the moderns) attribute agency to "natural" entities (physicalities) such as microbes, atoms and viruses, but do not accept the agency of interiorities (souls, cultural features) produced only by human beings and therefore different and dependant (discontinuity of interiorities). This contrasts with the animist view of the world where there is a continuum of interiorities (spirits, souls) between humans and non-humans, and an infinite diversity of physicalities (bodies). The agency of memes seems to require much more than a shift from one starting point to another.

An approach more open to emergence can be found in two papers dealing with collective attention. Sasahara and al. (2013) put the emphasis on events that occur on Twitter, as measured by a deviation from the probabilities distribution (Jensen-Shannon divergence) which helps to detect an enhanced popularity of some terms or hashtags. This detection of "collective attention" gives access to memetic phenomena (through retweets and number of solo tweets related to the same topic) more than virality (only RT). This is clearly visible in the comparison between 4 events (earthquake, World Cup, TV series, total lunar eclipse) for which the aggregation of RT looks very different.

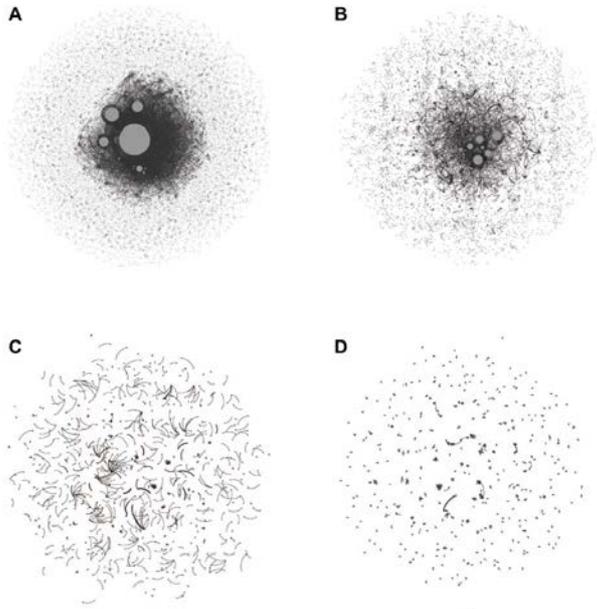

Figure 6. Visualization of RT networks during collective attention. (A) Tohoku-oki earthquake (# of nodes = 27,340, # of links = 27,709), (B) FIFA Women's World Cup 2011 (# of nodes = 9,277, # of links = 8,450), (C) Castle in the Sky (# of nodes = 1,183, # of links = 793), and (D) total lunar eclipse (# of nodes = 893, # of links = 553). The nodes represent users, which are connected if there is a RT with key terms related to the target event. In each figure, the node sizes are proportional to the number of retweeted tweets. Only the nodes with $k \geq 5$ retweeted tweets are shown for clarity.
doi:10.1371/journal.pone.0061823.g006

A similar approach based on the difference between events was used by Lehman and al. (2012). A sports event, a school shooting, the release of a blockbuster movie, and a political discourse display very different patterns of propagation on Twitter.

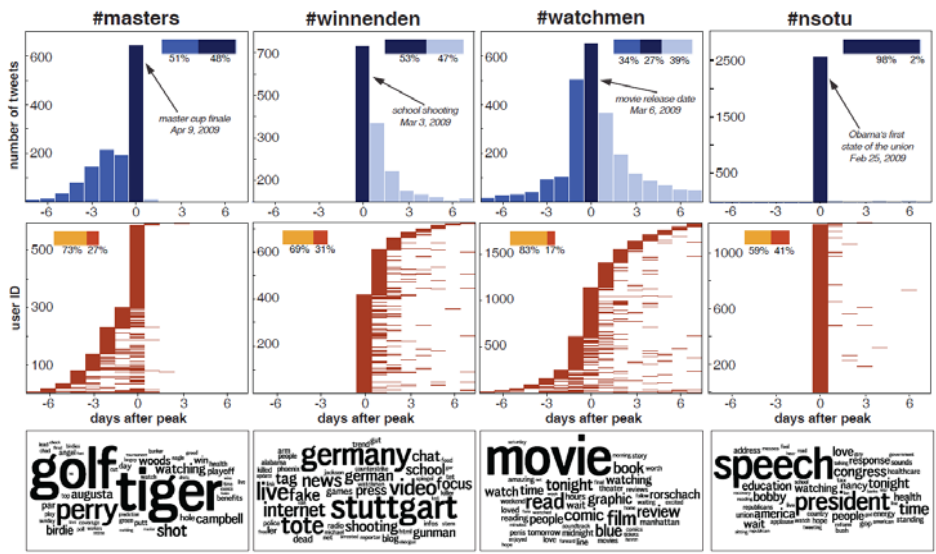

Figure 1: Activity associated with four hashtags that exhibit a popularity peak: daily activity over time (top row), individual user activity (middle) row, and word clouds of tweet content (bottom row).

However, as we can see, patterns are related to the topics and not to the semiotic features of the hashtags or tweets. The distinction we built between structure, market and emergence seems to be still valid within the study of Tweets: even though some papers try to capture the role played by the features of each tweet, they often fall back into the detection of patterns that are more related to statements about the structure of the world "out there". This has still to be tested and calls for a combination of expertise from other fields and of algorithms with different purposes.

### *Temporal dynamics in graphs*

One pre-condition to reach this goal is to understand better what propagation phenomena are, and especially how time can be integrated as an internal feature of the process and not only as the commonly shared reference. If we move from timelines to rhythms, for instance, we gain some access to the emergence of events for sure. This is why a recent focus of scientific attention on temporal dynamics in the graph research community is worth discussing. Gomez Rodriguez and al. (2014) developed a network inference algorithm to detect the underlying network among 179 million different instances of information in a one-year period. They made the point of a different clustering of nodes between recurrent topics and on-going news events that can "emerge and vanish in a matter of days". However, their aim was to connect the behaviour of individual nodes ("when a node learns about a piece of information, makes a decision, adopts a new behaviour or becomes infected with a disease") and the "underlying network connectivity". That is to say, they set out to make connections between an "Opinion" or "Market" model and a "Society" or "Structure" one, without questioning any agency of the entity that connects the nodes. This is more centrally addressed by Benzi and al. (2016) who use a "causal multilayer graph" (CMG) designed to follow the propagation of events in successive time steps". Significantly, their approach is a graph signal processing one, very far from any Natural Language Processing concern, although they do address textual content propagation issues. This radical move for a reduction of entities to signal processing seems to be more akin to traces data than to many other high-level analyses, and this is a frequent lesson from assemblages of social sciences and computer sciences (Ghitalla, 2016). The events become binary states of nodes that can be tracked in a multilayer graph where layers are time-steps, not only the neutral obvious reference but a feature of the nodes in the graph. The clustering of the propagation graph is not prearranged and the relevant number of clusters is not even stable but depends on a k-means algorithm. These precautions are interesting moves towards algorithmic choices in favour of emergence detection: no social entities pre-decided by social science or computer science rules, no pre-set level of clustering that is so often very useful to encapsulate opposition or asymmetries patterns. Of course, the use of terms such as "causality" for the sequence of events seems rather awkward or naïve for social scientists as for instance in "Granger causality". But the decision to track signal changes through time-steps seems a rather promising reduction move. This is powerful for virality but less so for memetics, which require an investigation at the finer-grained level of the content.

## 7. From traces to replications

To make the foundations of this third generation of social sciences more solid, scientific status should be given to these traces. As I have pointed out, the production of traces is directly dependent on platforms that generate their own analyses. However, just as polls do not only serve the media and censuses do not only serve administrations, as noted above, and according to Desrosières' analysis, it is necessary to base third-generation social sciences on a proposal of non-captive utilisation of these traces. To the register/survey and audience/opinion poll twosomes, we must add traces/X, where X is the place to be defined for the social sciences' use of traces.

Let us then talk about "replications". It exhumes a suggestive metaphor with earthquake vibrations (aftershock): it is possible to follow the seismographic replications that anticipate and follow the shock itself. The semantics of the term "replications" focuses attention on the waves rather than on the particles, and on conversations echoing the 'buzz' that obsesses brands and media, but that is never theorized. Other concepts could claim this spot like that of attention, influence, "issues", actor-network, memes or conversations. Replications manage to achieve the necessary shift vis-à-vis notions of actors, strategies and representations, all of which have their legitimacy in the context of other social sciences but do not allow for these circulating entities' **power to act** (that are the replications' agency) to be taken into account. We cannot say a priori what the size or status of these entities are because it is only the mass corpus investigations that allow us to identify them, when their replication emerges from the sensors we exploit, and certainly from platforms but according to *our* objectives. The principle of a sociology of replications relies on the need to follow the elements in order to detect waves, without knowing how they will join together to make a "whole" of variable geometry. The bias is "elementarist" but should above all not be atomistic (because the variable geometry remains a quality, as ANT has taught us) nor a move towards social physics, which seeks supposed transversal laws for all this fluctuation. The replication approach allows us to build an infinite combinatory, following extensions, propagations and repetitions, provided we remain focused on "issues" (as N. Marres enjoined us to do) that carry and enliven replications and in a very different way to "raw" traces. The subject of study is not so much the element, which can have a variety of attributes, nor solely the aggregates (what we tend to do with clustering from graph methods (Fouetillou 2008)), but rather the process of movement and aggregation or disaggregation at the time of a curves' bifurcation. In these curves it is then necessary to focus instead on the moments of emergence and not on the peaks that function as aggregates, as the first generation meme-tracker does. The object of this science of replications is the agency of the replications that spread and end up enveloping us – because people are actually traversed by ideas and ideas make us act, not the reverse, as Tarde clearly pointed out (1893). "The imitation rays first and then the beings, whose existence we infer from the transformation they undergo with the flow of imitation" (Latour, 2011). It is then possible to study the properties of these replications, to potentially compare their chances of survival or contamination. This is made possible by differences in their properties, which are always directly related to the "issues" that they convey. The replication approach is then a means of entering into a monadology (which radically differs from an atomistic vision) (Tarde, 1893). Traceability is however not given as such by platforms, and requires the production of tools and methods suited to a replication approach and not only to traces, as in the examples listed above. Leskovec and Kleinberg (2009) appear to have been the forerunners in this area with the proposal of their meme-tracker. They are indeed able to restore flows of all types, what they call streams and cascades, or replications. The development of traceability methods for replications should take the existing into account, by ensuring the pre-testing of corpora established for this purpose, even if this means losing the "realism". I started this work in 1987 with the monitoring of TV conversations and their transposition onto workplaces to make local public opinion (Boullier, 2004). I then proceeded to monitor the attributes of a photo from the Flickr database in the same way (Boullier and Crepel, 2013), along with transposable signs in a corpus of web sites linked to a region (Le Béchec and Boullier, 2014). In the first case, these are the attributes of the photo (e.g. crossed arms), which become tag attractors and thus connect accounts or photos that were, a priori, never connected according to these criteria. But I stopped at the very principle of this work without being able to conduct it empirically, on a sufficient scale. In the second case, the spread of the Breton flag on the Web became an indicator of a connection that we can compare with other entities constituting regional imagery but which failed to spread. From work done by hand on nearly 600 sites, by M. Le Béchec, it was possible to outline what the analysis of replications generated by this flag would be, and to observe that its semiotic properties were not extraneous to its capacity for circulation. Tags or icons are replications that can be followed, even if they have neither the explicit character of verbatim or expressions, as in the meme-tracker, nor their massiveness.

Certain approaches from these digitized corpora (not native to digital technology) can give an idea of the potential of such methods. Work done on the n-grams studied from Google Books (Michel et al., 2011) showed the evolution of the English language (the preterit of irregular verbs). Lev Manovich (2012), leader of a "cultural analytics" trend, created a base of over a million examples of manga to compare their most basic attributes, such as contrast, and produce unique insight into influences between trends. He used similar tools to conduct cultural comparisons between countries from millions of photos on Instagram or from Maidan Square in Kiev.

Potentially, all the traces that we have identified (such as likes, tweets, recommendations, etc.) may be the object of this type of monitoring. However, they require specific tracking tools that exist largely for Twitter only, although provided that a detailed review of these tools is done to ensure that they meet the specifications of a traceability of replications (not just traces for the sake of it or for the reactivity of brands).

## 8. The warnings of "digital methods"

As I have tried to demonstrate, the mere fact of addressing digital material does not pre-empt the kind of social sciences that will be delivered. All three approaches (society, opinion, replications or structure, market and emergence) can find their specific routes to make sense of these huge amounts of data. However, I emphasized the need to account for entities, scales, rhythms that are made accessible by digital networks: real-time and high-frequency ones. Even at this level, many methodological choices have to be made in order to achieve a "replications" analysis. This is the condition for not depending on the agenda of platforms and brands which are providing us with these traces data. Computer science as well as social science should listen more to the advice and warnings of experts of "digital methods".

The third generation of social sciences will not be able to do much apart from associating with the digital platforms and the brands to produce a science of traces that would then be treated as "replications". The traces produced are platform-dependent; we can hardly expect to modify them at the source. Nevertheless, it is possible to exploit the traces produced by the platforms by diverting them from the purpose for which they were designed (repurposing, Rogers, 2013) – the rule here being that we do not take any explanation at another level or another world into account, but that we might compare propagation speeds, rhythms, and possible transformations (e.g. contamination of other areas, etc.). The difference should be the ability to see the processes that have not yet been identified, either because of the limits of pre-digital technology or the targets adopted by previous generations of social sciences. As R. Rogers proposed in his pioneering work, this "repurposing" of traces will imply a fine tuning of the "query design", on Google or on any API. He argued this should rely on a well-defined hypothesis and not only follow the opportunity of inference offered by Big Data and machine learning technologies. Wikipedia represents a very unique set of data for scholars as the project Contropedia demonstrated, because this platform keeps track of all revision history, of debates and of controversies, while being able to generate trade-offs which are recorded as such. R. Rogers analysed politically controversial issues such as abortion and the Srebenica massacre ("labelled "fall", "genocide", or "drama", depending of the version). However, Wikipedia does not have any equivalent on the Web when it comes to traceability and openness. Rogers also redesigned Twitter's data as a "narrative machine" to give an untold account of the Iranian revolt in 2009, based on the 600,000 tweets produced as a corpus. In this case, the propagation effects can be demonstrated, as for the slogan *"Dégage !"* during the Arab spring.

N. Marres seems more concerned by the critics of scholars' overwhelming dependency on the platforms that deliver the data. The way she handles socio-political controversies as "issues" has proved very inspiring, in that she designs limits of validity to empirical researches (for instance, no general tracing of tweets- or weibos- or any other traces without delineating the arenas created by "issues"). She suggests that we should not attempt to purify the parameters of the media since they are an intrinsic part of the process. But she still considers that it is possible to revise them the way she did for the controversy about DPI (Deep Packet Inspection) at the 2012 WCIT (World Conference on International Telecommunications) in Dubai. Marres combined the most frequent Twitter hashtags with the terms used by experts in their papers, in order to redesign the queries on this Twitter corpus more independently from the platform. This should be part of the conventions that have to be built, in which the sources are crawled and considered as legitimate for scientific research, provided that the media framing effect is not discarded but is part of the research. This is exactly how scholars deal with the conditions of production of polls or any surveys in a non-natively digital world. These warnings are essential to keep in mind if we want to avoid the massive trend towards "algorithmic positivism" which is prevalent in many computer science papers. Yes, indeed, the social sciences must keep on questioning the categories, the classification processes and the concepts even before the harvesting and computing processes are launched. Yes, it will slow down the reactivity of platforms and the transfer of technology to social listening companies. But it will guarantee that the conventional wisdom encapsulated in the algorithms comes under scrutiny and is not taken for granted. When L. Lessig (1999) told us that "code is law" and that technical architectures are policies, this should have been addressed to scientific activity as well, for at the moment it depends so much on the computing architectures that are built for any research program.

# 8. How can the science of replications manage the Big Data and Machine Learning shifts

The challenge is compelling for the social sciences: what can they learn from Big Data and Machine Learning in order to survive and to reinvent themselves? The call for reaction instead of reflexivity is supported by decision-makers from all social worlds: private companies, public ones and even politicians, not to mention the media that are in trouble when accounting for social science research programs, more than when they tell another story of a "solution" that machines discovered, without any theoretical framework. "Solutionism" (Morozov, 2013) and "The End of Theory" (Anderson, 2008) work together to put social sciences under pressure to justify their social relevance.

This pressure not only requires that one explore the new entities made visible on digital networks, it also needs a strong stance about what is worth appropriating from Big Data and Machine Learning principles. I will separate these two trends, for contrary to widespread belief, they are not always combined. A lot of computer sciences research is still governed by a "model to data" approach, even though it handles Big Data. A "data to model" orientation – the one that Machine Learning is empowering so well, requires many conditions that are not often met.

## *Big Data or Fast Data?*

Big Data's quality criteria are often summarized in 3Vs: Volume, Variety and Velocity. The affinity with the requirements of the social sciences, exhaustiveness and representativeness, is quite striking.

### Volume and exhaustiveness

Volume corresponds to the need for exhaustiveness, resulting in a somewhat limited mode, because nobody and nothing can define the boundaries of the data collected, especially on the Web. IIt would therefore be advisable to fix an equivalent volume that is closer to traditional exhaustiveness requirements, without being able to respect those requirements when we deal with the Web. We clearly need to accept the death of exhaustiveness but that does not mean dispensing with all conventional frameworks for social science approaches when dealing with digital traces.

### Variety and representativeness

The second criterion, variety, is also a form of transcription for the representativeness requirements that allowed all the social sciences to proceed with inquiries and surveys based on sampling. Again, the test is a loose version of representativeness, which assumes that we accept a *sufficient* level of variety. For the third generation of social sciences, what this variety would be remains to be defined. The establishment of a set of sources (sourcing) in studies of the Web should then adhere to some criteria, specific to digital methods and to the field of study. My work on opinion mining (Boullier and Lohard, 2012) has led me to consider that no description of social-society, social-opinion or social-replications can be produced in general on digital networks. The proliferation of traces paradoxically makes it impossible to claim to refer to a "whole" posited a priori or established a posteriori. The social sciences must agree to deal solely with "issues" (Marres 2007, Marres and Weltewrede 2013), or on focal points of attention, or on "oriented and situated engagements" (Hannerz, 1983), for which traces can be kept digitally, traces that are specific to each outcome or each engagement. This significantly reduces Big Data's claims of a totalizing scope, but it makes a certain equivalence of representativeness and exhaustiveness possible.

**Velocity and traceability**

The last criterion, velocity, hardly finds a parallel in the first and second generations of the social sciences. Indeed, these dynamic processes were neither their *forte* nor their concern. It was essential to seek primarily to represent the positions at a given point in time, *t*, to show the strength of "society" that moulds the diversity of individual behaviours or to show how public opinion is structured beyond singular expressions obtained in surveys. Through a longitudinal follow-up of the same populations or by reusing the same questionnaire it is of course possible to deliver the equivalent to dynamism, but without ever being able to track the mediations that would produce these changes. Velocity seems outside the scope of conventional approaches.[9] However, a branch of Web science has also seized upon the issue of velocity in its own way by exploiting the meme traces that spread on the Web. It is very significant that Kleinberg, the very man who had exported scientometric methods to the study of Web topology – methods that were taken-up by Google –, has for several years[10] (Kleinberg, 2002) taken memes into consideration. Memes seem promising to us, provided that we also track the transformations-translations of these memes in different environments and not only the virality.

It therefore becomes possible to find an equivalent for the velocity of Big Data: *traceability*. This becomes the essential quality criterion for entities that can be studied. Some felicity conditions must however be fulfilled to achieve this:

(a) The traces in question must have sufficient continuity to make it even possible to say that this is the same process.

(b) The traces in question must allow the tracking of heterogeneous associations, that is to say, a sufficient power of connectivity. For this reason, traces with overly specific formats to a little-known platform can give way neither to extension nor to tracing.

(c) The monitoring of the traces in question should enable precise dating of all the events, all the changes and all the associations. The timelines are equivalent here to other conventions, such as the cardinal points in topography or wealth inequality in the social sciences of the first generation. These conditions make it possible to tilt the social sciences into tracking things that are neither individuals nor groups, neither society nor opinion. Digital has dissolved all certainty of their statutes, which had already been called into question. The spread of these elements, which we will have to qualify above and beyond traces, becomes the third social science generation's object of study because it is the properties of these entities that allow them to create small differences and, from there, to circulate and affect individuals and groups, societies and public opinion. But no structure or propagation law is to be sought in the way social physics is looking for them (Pentland, 2014), for each medium (platform, problem, format, condition) is specific.

## *Machine Learning versus social theory?*

These are the conditions in which Big Data qualities may become useful and acceptable for social sciences and specifically for sciences of replications. However, it says nothing about the process of these data and the major shift in scientific methodology brought in by Machine

---

[9] I am not including here the approaches of social physics, ethology or social epidemiology that have produced social models without reference to the traditions of the social sciences.

[10] J. Kleinberg shares a detection of "topics" that cause "bursts" in the email flow, intended to help to structure them and not to follow them as such. But he made the connection with the contemporary problems in the narrative, such as those treated by Genette.

Learning. In the first generation of Artificial Intelligence, ontologies were prevalent, in a "model-to-data" approach very often limited to a well-known and well-controlled domain or activity. And social science had the opportunity to undermine the reputation of these models compared to the "real" world, the "real" organization or the "real" way language works. Machine Learning methods put an end to that game. What the Volume and the Variety of Big Data bring to Machine Learning is the opportunity to test the correlations in an iterative way at almost no cost. But this technical change brings an amplification of a more important epistemic trend, inductive reasoning, which seems to transform the hypothetico-deductive mode of doing science in an outdated practice. The entire framework relies on probabilities and is thus dependent on the quality of the dataset on which the computation will be made, which explains the decisive change brought about by Big Data, without masking the risk in extension of models from old data sets to predictions in unknown conditions. But this is precisely what probabilities are all about. Two algorithmic moves are critical for the processing of such amounts of data:

- **Classification**, a very well-known requirement of almost all programming activity. Using Machine Learning methods there is no need to preset a model of classes; only the learning model is required, which will let the program look for the right classification algorithm within a library of algorithms. Some models became famous for limiting the number of examples needed for the learning process to take place, for instance the Vapnik-Chervonenkis's theory. This means an important shift in the modeling process: the model formalizes not the "world" "out there" (the domain, the population used as a reference) but the method to learn from examples. This is a second degree modeling which cannot be challenged by any "realistic" statement. Only the results can be tested since these models are supposed to become predictions. The clusters that are obtained by these methods should not be obvious ones and should help learn something new about the assemblages of the social. Unfortunately, a large part of the results are disappointing because they demonstrate tautologies or obvious and well-known features of social behaviors.
- **Similarity** is another key process in these approaches, and is also a well-known issue in computer science. In order to produce classifications, similarity detection methods have to be designed. Much of the similarity is still delivered by humans, be they experts or crowdsourcing platforms, in linguistic or image corpora. This is the way theory is short-circuited: no need for a theory of language or a semiotic one for image recognition. A good sampling method is still required to expand the knowledge obtained from humans, and Machine Learning is quite good at that sampling procedure. The "whole" of a discourse or image is broken down into so many features that similarity can be obtained by very non-intuitive ways, as in facial recognition algorithms. The "whole" of a dataset is not a precondition in sampling-based algorithms, or for variational algorithms which can approximate precisely enough the similarities by latent analysis methods.

What should be challenging for the social sciences is this ability of Machine Learning to adopt an agnostic attitude towards all kinds of categorization on which hundreds of social science papers have been written. Categorization is a key process in all the social sciences but refers to theoretical models, to a carefully designed set of concepts and usually cannot really be put to the test because quantification methods are designed along with these basic categories, even in censuses. Moving to classification with machine learning methods, even supervised ones, means either adopting some kind of "algorithmic positivism", because learning algorithms themselves involves a conventional wisdom bias, or else entering into a reinvention cycle of the traditional methods. Third-generation social sciences should accept the challenge but with a clear theoretical ambition. As I have stressed in this paper, traces data can be addressed from three different viewpoints, irrespective of the sophistication of the Machine Learning processes. The focus on replications opens up some opportunities not to stick to the classical social determinations, but it does not mean that any correlation provided by the algorithms would work. The composition (Latour, 2010) between the suggestions provided by the machines and the scientific awareness of the social construction of categories and knowledge should be designed in procedures that are still to be tested. The prospect of an "interpretable Machine Learning"( Rudin, 2014) or even of an "interpretable-by-design"[11] kind of machine learning gives us some clues for inspiring new procedures in machine learning algorithms in which some room can be made for interpretation and discussion provided by the social sciences. The only fact that these interpretations do exist in the way the calculation is done by computer scientists (classification decisions, editorial labelling of classes and so on), even though only the final black box is delivered to the public, advocates the feasibility of these new processes open to interpretation from other fields.

---

[11] Daniel Gatica-Perez, 2016, personal communication.

# Conclusion

The hype about social data, data mining and Big Data cannot provide solid ground for the new conventions which have to be designed for the sciences of replications. We must remember the huge likelihood that all these traces, which could still be connected to personal data, will not be accessible in the same way in a few years. The success of Adblock Plus that blocks cookies and other intrusive ads is constantly growing (200 million downloads, 40% installation on Firefox). Generalized encryption will become a necessity given the inability of platforms and intelligence agencies to regulate their own predatory activities with personal data. This is why taking traces into account, even on the surface of networks, and without any link to structured and meaningful socio-demographic data, provides a solid foundation for the social sciences. The contrast is stark with all those who continue to wish to apply models of society and opinion to a universe that is available to them only because of a very temporary laxness. Working on the surface of these traces, in isolation from personal data, also reduces the ethical contradictions in which the social sciences of society and public opinion, that want to exploit these sources, find themselves bound up.

By advocating this "surface work", I adopt a radical empiricism approach (James, 1912) consisting in following the digital traces at their face value, looking for where they come from (traces produced by platforms) and for how they transform and are transformed by the very milieu in which they live, and refusing to reduce them to an equivalent of any other phenomena in society or in opinion. This constraint strongly limits the power of "explanation" of these traces but it is consistent with other approaches that consider the "platform bias" as a component of the analysis (Marres). These high-frequency propagation phenomena cannot divide traces (elements) and platforms (milieu) and must account for the distribution of agency which is never settled as an a priori. Following them allows us to compute these imitation processes (that include opposition and invention, as Tarde said). This approach does not contest the legitimacy of other analyses of the social as long-term social structures, or as mid-range opinion movements. Furthermore, in the tradition of Science and Technology Studies, it claims to help the three viewpoints of the social to accept each specificity and value, without disqualifying any of them or pretending to assemble all in a holistic view free of any mediation. It should help the classical social sciences to seriously consider the extension of the social to new entities: the replications that could not be captured before the digital era. At the same time I would like to avoid two traps: using digital traces for the documentation of the "social-as-usual" (society or opinion), or reducing them to a set of clever tricks of method. The social sciences should become able to play their full role in the digital world dominated by platforms and brands. Our responsibility is to build the conventions for a new layer of social sciences that maintain the requirements of scientific reflexivity.

**Acknowledgements**
This paper benefited from the comments and critics of E. Didier, G. Bowker, B. Latour, N. Mayer, D. Cardon, N. Marres, G. Fouetillou, S. Parasie, R. Rogers, F. Cochoy, M. Wievorka, F. Thibault, B. Benbouzid, Y. Citton et M. Legrand, and from the collective work done during a seminar in 2015 about Third Generation social sciences at the Fondation Maison des Sciences de l'Homme in Paris. The English has been revised by Jim Hogan and Lizz Lilbrecht.